\documentclass[aps,showpacs,prl,twocolumn,footinbib]{revtex4-1}
\usepackage{epsfig}
\usepackage{amsfonts}
\usepackage{amsmath}
\usepackage{bm}
\usepackage{xcolor}
\usepackage{newtxtext,newtxmath}

\usepackage[utf8]{inputenc}
\usepackage[colorlinks,bookmarks=false,citecolor=blue,linkcolor=blue,urlcolor=blue]{hyperref}
\usepackage{soul}
\usepackage{stackengine} %Panel labels 

\usepackage{changepage}

\newcommand{\ket}[1]{\left| #1 \right>}

\newcommand{\av}[1]{\left< #1 \right>}

\newcommand{\tr}{\mathrm{Tr}}

\newcommand{\Sx}{S^x}
\newcommand{\Sy}{S^y}

\newcommand{\sxk}{{\sigma^x_k}}
\newcommand{\syk}{{\sigma^y_k}}
\newcommand{\szk}{{\sigma^z_k}}
\newcommand{\smmk}{{\sigma^-_k}}
\newcommand{\sppk}{{\sigma^+_k}}

\newcommand{\set}[1]{\left\{ #1 \right\}}
\newcommand{\sL}{\mathcal{L}}
\newcommand{\sD}{\mathcal{D}}
\newcommand{\re}{\mathrm{Re}}

\newcommand{\abs}[1]{\left| #1 \right|}

\newcommand{\N}{\mathbb{N}}
\newcommand{\Z}{\mathbb{Z}}

\newcommand{\rot}{{(R)}}
\newcommand{\zer}{{(0)}}

\newcommand{\be}{\begin{equation}}
\newcommand{\ee}{\end{equation}}

\newcommand{\rmd}{{\rm{d}}}

\newcommand{\rme}[1]{{\rm{e}}^{#1}}

\begin{document}

\title{Discrete time crystals in the absence of manifest symmetries or disorder in open quantum systems}
\author{F. M. Gambetta, F. Carollo, M. Marcuzzi, J. P. Garrahan, and I. Lesanovsky}
\affiliation{School of Physics and Astronomy, University of Nottingham, Nottingham, NG7 2RD, United Kingdom and Centre for the Mathematics and Theoretical Physics of Quantum Non-equilibrium Systems, University of Nottingham, Nottingham NG7 2RD, UK}

\date{\today}

\begin{abstract}
We establish a link between metastability and a discrete time-crystalline phase in a periodically driven open quantum system.  The mechanism we highlight requires neither the system to display any microscopic symmetry nor the presence of disorder, but relies instead on the emergence of a metastable regime. We investigate this in detail in an open quantum spin system, which is a canonical model for the exploration of collective phenomena in strongly interacting dissipative Rydberg gases. Here, a semi-classical approach reveals the emergence of a robust discrete time-crystalline phase in the thermodynamic limit in which metastability, dissipation, and inter-particle interactions play a crucial role. We perform numerical simulations in order to investigate the dependence on the range of interactions, from all-to-all to short ranged, and the scaling with system size of the lifetime of the time crystal.
\end{abstract}

\maketitle

\textit{Introduction ---} 
Time crystals have been introduced as an intriguing non-equilibrium phase of matter~\cite{Sacha:2018} in which time-translation symmetry is spontaneously broken~\cite{Wilczek:2012,Shapere:2012,Wilczek:2013, Li:2012}. The first proposal by Wilczek~\cite{Wilczek:2012} has triggered an intense debate~\cite{Bruno:2012comment,Bruno:2013comment} which culminated in a series of counter-examples and no-go theorems~\cite{Bruno:2013,Watanabe:2015,Nozieres2013} concerning their realization in equilibrium.
The search for time crystals then turned to non-equilibrium systems. In this context, periodically-driven (``Floquet'') quantum systems~\cite{Shirley:1965,Sambe:1973,Grifoni:1998} have played a major role. Indeed, it has been shown that a new phase of matter, called \emph{discrete} time crystal (DTC), may emerge under periodic driving~\cite{Sacha:2015,Khemani:2016prl,Else:2016prl,vonKeyserlingk:2016,Khemani:2017,Russomanno:2017,Yao:2017,Huang:2018,Choi:2017,Zhang:2017,Ho:2017,Moessner:2017,Else:2017prx,Rovny:2018prl,Rovny:2018,Sacha:2018,Pal:2018,Yu:2018,Barfknecht:2018}. In such cases, with $T$ being the period of the driving, the \emph{discrete} time-translation invariance under $t\to t+T$ may be spontaneously broken, with observables exhibiting subharmonic responses, i.e.~oscillating with a period which is an integer multiple of $T$. \\
Several efforts have been directed to the study of DTCs in non-dissipative quantum systems~\cite{Khemani:2016prl,Else:2016prl,vonKeyserlingk:2016,Khemani:2017,Russomanno:2017,Yao:2017,Ho:2017,Moessner:2017,Huang:2018,Pal:2018,Yu:2018,Barfknecht:2018}. Here, since in principle the driving would eventually heat the system to infinite temperature thereby destroying the crystalline order, the presence of disorder and localization is often seen as an essential requirement to prevent this from happening and to obtain a DTC that survives asymptotically~\cite{Lazarides:2014, Ponte:2015,Ponte:2015prl,DAlessio:2014,Lazarides:2014,Kim:2014,Else:2016prl,Khemani:2016prl,vonKeyserlingk:2016,Lazarides:2015prl,Choi:2017,Zhang:2017,Moessner:2017}. Alternatively, DTC order can be sought as a transient feature emerging in a prethermal regime~\cite{Mori:2016prl,Kuwahara:2016,Zeng:2017,Else:2017prx,Rovny:2018prl,Pal:2018,Haldar:2018}. A relevant issue concerning the realization of DTCs has been their fragility upon the coupling to an external environment~\cite{Lazarides:2017, Zhang:2017,Choi:2017}. Nonetheless, an interesting approach has turned this perspective around showing that appropriately engineered dissipation can instead represent a resource for harnessing and tuning the properties of quantum systems~\cite{Verstraete:2009,Beau:2017}. This has motivated a recent interest in the possible emergence of time crystals in dissipative quantum systems~\cite{Nakatsugawa:2017,Gong:2018,Iemini:2017,Else:2017prx,Tucker:2018,Wang:2018,OSullivan:2018}.

\begin{figure}
	\centering
	\includegraphics[width=\columnwidth]{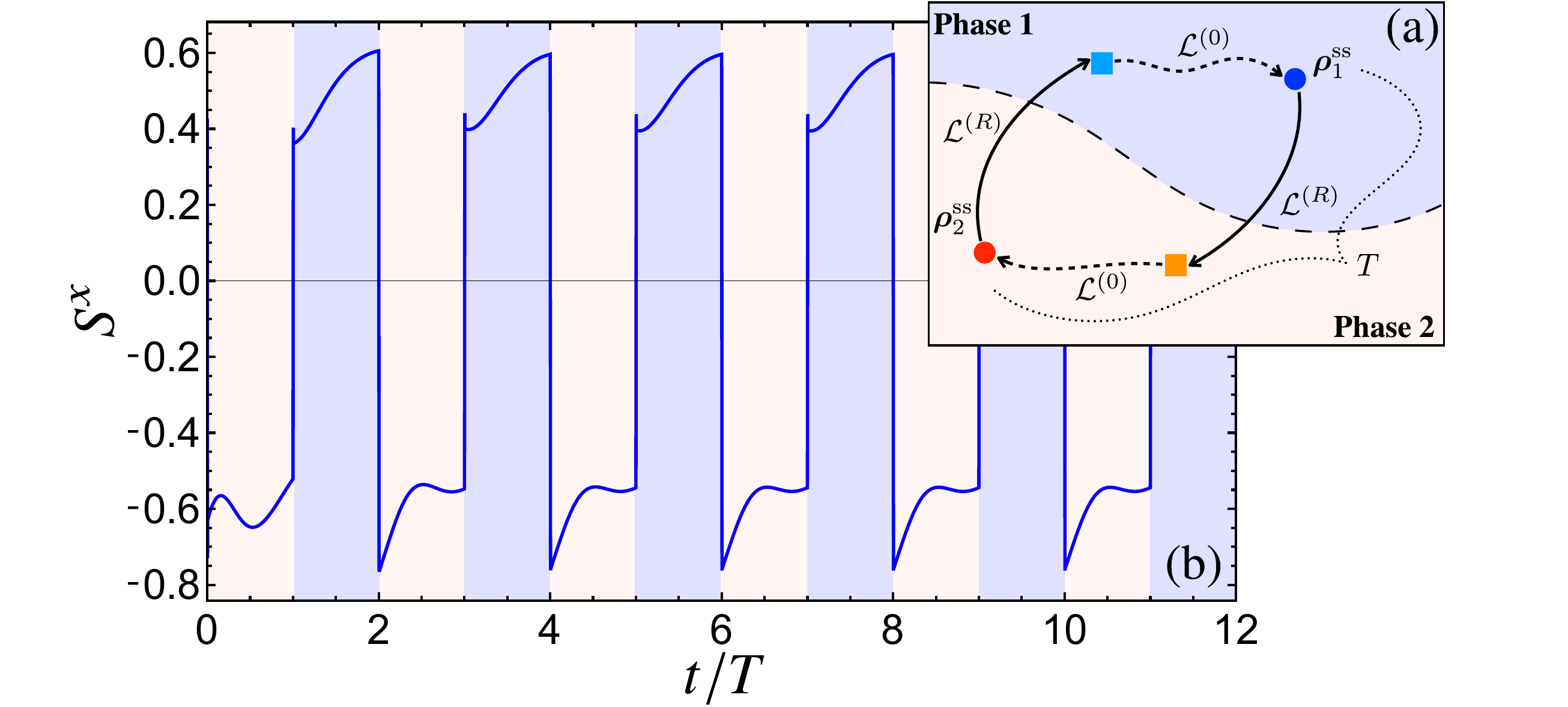}
	\caption{DTC in a metastable open quantum system. (a): The phase space of the system is divided into two basins of attraction (bright and dark areas), associated with two stationary states, $ \rho_1^\mathrm{ss} $ (blue dot) and $ \rho_2^\mathrm{ss} $ (red dot), respectively. The combination of an appropriate transformation ($ \mathcal{L}^\rot $) and dissipative dynamics ($ \mathcal{L}^\zer $) during a time interval $ T $ maps one of the stationary state into the other one. The repetition of this composed transformation makes the state oscillate with a doubled period $ 2T $. (b): Time-evolution of a typical observable in the DTC phase. Here, we show the expectation value of $ S^x(t) $ as a function of time $ t $ for a dissipative Rydberg gas (see text for details). The period of oscillations is twice the one of the driving (with shaded areas corresponding to different basins of attraction). Here, $ \Omega^\zer_x=0.7\ \Gamma$, $ V=12\ \Gamma$, $ \Delta^\zer=-3.5\ \Gamma$, $ T=2\ \Gamma^{-1} $ and $ t_R=10^{-2}\ \Gamma^{-1} $. }
	\label{fig:DTC-MF}
\end{figure}

In this work we establish a link between metastability in open quantum systems~\cite{Macieszczak:2016,Rose:2016} and DTCs. This provides a simple and generic mechanism (see Fig.~\ref{fig:DTC-MF}) for the emergence of a DTC under periodic driving, which does not hinge upon the presence of either disorder or of any manifest symmetry of the generator of the time evolution. To illustrate this mechanism, we employ an example taken from the physics of dissipative Rydberg gases \cite{Schwarzkopf:2011prl,Low:2012,Saffman:2010,Malossi:2014prl,Lee:2012,Carr:2013,Hofmann:2013prl,Hu:2013,Marcuzzi:2014,Sibalic:2016,Letscher:2017,Helmrich:2018}.  This system displays a stationary-state phase transition in  sufficiently large dimensions~\cite{Weimer:2015,Weimer:2015PRL,Sibalic:2016}. The concomitant closing of the spectral gap leads to metastability. We discuss in detail a protocol for achieving a DTC and investigate its stability as well as its lifetime.

\textit{DTCs from metastability ---} 
We consider a general Markovian open quantum systems with $N$ degrees of freedom (e.g., spins) whose dynamics is governed by the quantum master equation (QME) $ \partial_t \rho=\sL^\zer[\rho] $~\cite{BreuerPetruccione}, with $ \sL^\zer[\rho]=-i[H^\zer,\rho]+\mathcal{D}[\rho] $. Here, $ H^\zer $ is the system Hamiltonian while $ \sD $ describes Markovian dissipation. We denote the eigenvalues of the ``superoperator'' $ \sL^\zer $ by $ \set{\lambda_k,\ k=1,2,...} $ and order them by decreasing real part, i.e.\ $ \re(\lambda_k)\geq\re(\lambda_{k+1}) $. The (complete) positivity and trace-preserving properties of $ \sL^\zer $ guarantee that $ \lambda_1=0 $. Its associated right eigenmatrix $\rho^\mathrm{ss}$ represents the stationary state of the dynamics, i.e.~$\sL^\zer[\rho^\mathrm{ss}] = 0$~\cite{BreuerPetruccione}, which we assume to be unique at any finite size $N < \infty$. In the following, we focus on systems with vanishing gap for $N \to \infty$, displaying metastable behavior~\cite{Macieszczak:2016,Rose:2016}. Specifically, we require that, for some choice of the parameters of $\sL^\zer$, $\lambda_2\in\mathbb{R}$ and $\lambda_2 \to 0$ while $\liminf_{N \to \infty} \abs{\re(\lambda_3)}  > 0 $. This leads to a separation of timescales for large $N$. Indeed, defining $\tau_m = 1/\abs{\re(\lambda_m)}$, one can distinguish three different regimes: For $t \lesssim \tau_3$ there is a transient dynamics strongly depending on the initial state. For $t \gtrsim \tau_2$ the system instead approaches stationarity and its state converges to $\rho^\mathrm{ss}$. Under our assumptions, one can find a third time-frame, $\tau_3 \ll t \ll \tau_2$, which defines a so-called \emph{metastable regime}: Here, the dynamics can be effectively reduced to a space spanned by the eigenspaces of $\lambda_1$ and $\lambda_2$. Denoting by $R_2$ ($L_2$) the right (left) eigenmatrix of $ \sL^\zer $ corresponding to $\lambda_2$ this means that 
$\rho(t)= e^{\sL^\zer t}[\rho(0)]\approx\rho^\mathrm{ss}+ c_2 e^{\lambda_2 t} R_2 $,
with $ c_2=\tr[\rho(0)L_2] / \tr[L_2 R_2]$ the component of the initial state over $R_2$~\cite{Macieszczak:2016,Rose:2016}. The dynamics in the r.h.s.~takes place in this reduced space and it can be described in terms of classical jumps between the two extreme metastable states (eMSs) $\tilde{\rho}_1=\rho^\mathrm{ss}+c_2^{\mathrm{max}}R_2$ and $\tilde{\rho}_2=\rho^\mathrm{ss}+c_2^{\mathrm{min}}R_2$, 
with $ c_2^{\mathrm{max}} $ ($ c_2^{\mathrm{min}} $) the maximum (minimum) eigenvalue of $ L_2 $~\cite{Macieszczak:2016,Rose:2016}. In the thermodynamic limit ($N \to \infty$) the gap closes ($\lambda_2 \to 0$), determining a phase transition between two phases characterized by the properties of the two eMSs. At the transition point, the system becomes bistable and the two phases coexist on equal terms. Individual quantum trajectories will asymptotically approach either one or the other eMS, identifying the corresponding basin of attraction (BoA). Importantly, on timescales $ t\ll \tau_2 $ (and $N$ large enough), the system tends to behave as if it were in a bistable regime~\cite{Macieszczak:2016,Rose:2016}. The eMSs can therefore be approximately regarded as two effective stationary states.

As sketched in Fig.~\ref{fig:DTC-MF}(a), this phenomenology can be exploited to engineer a DTC. The key step consists of identifying a second dynamics, generated e.g.~by a Lindbladian $\sL^\rot $, which maps $\tilde{\rho}_1 $ to the BoA of $\tilde{\rho}_2$ and vice versa in a given time $t_R$. Fixing a period $T > t_R$ such that $T - t_R \gg \tau_3$, one can define the dynamics via
\begin{equation}\label{eq:Hdriving}
\sL(t)=\begin{cases}
%\sL^\zer &\text{for } t<0\\
\sL^\rot &\text{for } mT\leq t\leq mT+t_R\\
\sL^\zer &\text{for } mT+t_R<t< (m+1)T
\end{cases},
\end{equation}
with $ m \in \N$. For simplicity, we assume that the system  starts from one of the two eMSs (say, $\tilde{\rho}_1 $). For $t \geq 0$ the dynamics is clearly $T$-periodic, but the state of the system will instead evolve with doubled period $2T$, which is the hallmark of a DTC. The underlying mechanism can be understood in a pictorial way from Fig.~\ref{fig:DTC-MF}(a): By assumption, applying $\sL^\rot$ to $\tilde{\rho}_1$ for $t_R$ will bring the system into the BoA of $\tilde{\rho}_2$. The subsequent action of $\sL^\zer$ for a time $\gg \tau_3$ will bring the system to its metastable regime and therefore close to $\tilde{\rho}_2$ after the first driving period $T$. The second application of $\sL^\rot$ will then displace the state into the BoA of $\tilde{\rho}_1$ and the second instance of $\sL^\zer$ will bring it back (close to) $\tilde{\rho}_1$, closing the cycle at time $2T$. The repetition of these four steps will then reproduce the same physics, leading indeed to a $2T$-periodic dynamics and to the emergence of DTC order.

\textit{Dissipative Rydberg model ---} To test the general mechanism outlined above, we use a spin model for a dissipative Rydberg gas \cite{Low:2012,Gallagher:2005,Saffman:2010}, which we briefly introduce here~\cite{Note2}. It consists of $ N $ atoms on a lattice  whose ground state is coupled to a high-energy (Rydberg) level by a laser with Rabi frequency $ \Omega_x $ and detuning $ \Delta $. Two atoms populating simultaneously the Rydberg level feature a strong and long range dipole-dipole interaction. Employing an effective spin$ -\frac{1}{2} $ description and denoting with $ \ket{\downarrow} $ ($ \ket{\uparrow} $) the ground (Rydberg) state, the Hamiltonian of the model~\cite{Lee:2011,Lee:2012,Ates:2012,Low:2012,Marcuzzi:2014}, in the rotating-wave approximation~\cite{Barnett:2005}, is
$H^{(0)}=\sum_{k=1}^{N}\big[\Omega_x^{(0)} \sxk+\Delta^{(0)}  n_k\big]+\sum_{k\neq p}^N V_{kp}n_k n_p$. Here, $k$ and $p$ are site indices, $\sigma_k^{\mu}\ (\mu = x,y,z)$ denote the Pauli matrices acting on the $k-$th spin, $ n_k=(\mathbb{I}_k+\szk)/2 $ is the Rydberg number operator, and $V_{kp}$  describes two-body interactions between the $ k- $th and $ p- $th atoms. The finite lifetime of the Rydberg level, due to spontaneous emission at rate $\Gamma$, is introduced via a dissipative term $\sD[\rho]= \Gamma \sum_{k=1}^{N}\left[\smmk \rho \sppk- \frac{1}{2}\left\{\sppk \smmk ,\rho\right\}\right] $, with $\sigma_k^{\pm} = (\sxk-i \syk)/2$. The dissipative Rydberg model, whose density matrix $ \rho $ obeys the QME $ \partial_t\rho=\mathcal{L}^{(0)}[\rho]=-i[H^{(0)},\rho]+\sD[\rho] $~\cite{BreuerPetruccione}, displays a bistable behavior at the mean-field level \cite{Lee:2011,Lee:2012,Hu:2013,Marcuzzi:2014}. This regime is associated with the coexistence of two phases of an underlying first-order phase transition \cite{Weimer:2015PRL,Sibalic:2016}, which has also been observed experimentally \cite{Carr:2013}. The uniform mean-field equations of motion~\cite{Lee:2011,Lee:2012,Marcuzzi:2014} -- see Eq.~\eqref{eq:MFEoM} below and Ref.~\cite{Note2} -- are defined in terms of the expectation values $S^\mu = \av{\sigma_k^\mu}$  and $n = \av{n_k}$, where the site index $k$ is dropped assuming translational invariance.
\begin{figure*}
	\centering
	\includegraphics[width=\textwidth]{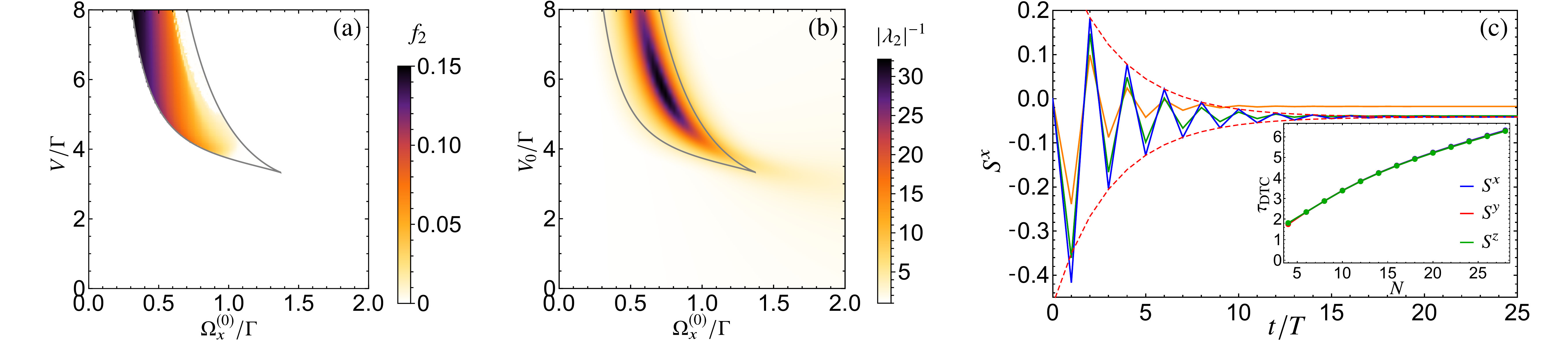}
	\caption{(a): Mean-field phase diagram of the normalized Fourier component $ f_2 $, evaluated numerically over $ K=30 $ periods, as a function of $ \Omega^{(0)}_x $ and $ V $. The region delimited by the two gray lines corresponds to the bistable regime, with the dark zone associated with the DTC phase and the bright one to the normal phase. The result suggests that $f_2$ vanishes continuously at the right boundary, whereas it undergoes a discontinuous jump at the left one. (b): Density plot of $ |\lambda_2|^{-1} $ as a function of $ \Omega^{(0)}_x $ and $ V_0 $ for the Rydberg fully connected model. Lines denote the corresponding mean-field bistability region for comparison. (c): Stroboscopic time evolution of $ S^x(t)$ for a system of $ N=28 $ (blue), $ N=20 $ (green), $ N=12 $ (orange) particles and $ \Omega^\zer_x=0.7\ \Gamma $, $ V_0=6\ \Gamma $. At $ t=0 $ the system is in the state with all spins pointing down in the $z$ direction and the transformation parameters are obtained from Eq.~\eqref{eq:rotation} with $ V=V_0 $. Red dashed lines represent a fit of the stroboscopic data for $ N=28 $ with the functions $ g_{\pm}(t)=a\pm b e^{- t/\tau_{\mathrm{DTC}}} $. Inset: lifetime of the DTC oscillations, $ \tau_{\mathrm{DTC}} $, extracted from the stroboscopic dynamics of $ S^{x/y/z}(t) $ as a function of $ N $. In this range of $ N $ its functional behavior is well captured by the power law $ \tau_{\mathrm{DTC}}(N)\sim N^{\alpha} $, with $ \alpha\approx 0.5 $. In all panels, $ \Delta^\zer=-3.5\ \Gamma $, $ T=2\ \Gamma^{-1} $, and $ t_R=10^{-2}\ \Gamma^{-1} $.
	}
	\label{fig:DTC-PD}
\end{figure*} 
In an extended region of parameter space, a slice of which (at $\Delta^\zer$ fixed) is enclosed by the gray contour in Fig.~\ref{fig:DTC-PD}(a), these equations feature two stable asymptotic solutions, $ \bm{M}^{\mathrm{ss}}_i=(S^x_i,S^y_i,S^z_i)\ (i=1,2) $, which will play the role of the eMSs $ \tilde{\rho}_1 $ and $ \tilde{\rho}_2 $. Outside this region, the stationary values are unique. In any given slice of parameter space, the bistable mean-field region (if present) is delimited by two spinodal lines coalescing into a critical point~\cite{Note2,Marcuzzi:2014}. A first-order line, passing through the critical point, is present within the latter, where $\bm{M}^{\mathrm{ss}}_1$ and $\bm{M}^{\mathrm{ss}}_2$ can be related via an emergent $\Z_2$ symmetry \cite{Marcuzzi:2014,Note1}. The exploitation of an emergent, rather than manifest, symmetry to engineer a DTC differs from the majority of earlier works, which assumed the latter as a necessary requirement~\cite{vonKeyserlingk:2016,Khemani:2017,Khemani:2016prl,Zhang:2017,Russomanno:2017,Gong:2018,Yu:2018,Pal:2018}. Note that a seeming exception to this statement can be found in Ref.~\cite{Wang:2018}. Here, period doubling in two-time correlation functions of a dissipative and periodically-driven asymmetric double-well potential has been observed. However, this behavior does not survive, for general initial states, in single-time observables.

\textit{Implementation of the DTC protocol ---}
To implement the time-dependent protocol described in Eq.~\eqref{eq:Hdriving}, we make the Rydberg model Hamiltonian explicitly time-dependent: $ H(t)=\sum_{k=1}^{N}[\Omega_x(t) \sxk+\Omega_y(t) \syk+\Delta  n_k(t)]+\sum_{k\neq p}^N V_{kp}n_k n_p $, where the parameters $ \mathbf{\Omega}(t)\equiv\left\{\Omega_x(t),\Omega_y(t),\Delta(t)\right\} $, with (complex) Rabi frequency $ \Omega_x(t)+i\Omega_y(t) $ and detuning $ \Delta(t) $, are $ T- $periodic functions:
\begin{equation}\label{eq:omegadriving}
\mathbf{\Omega}(t)=\begin{cases}
%\sL^\zer &\text{for } t<0\\
\mathbf{\Omega}^\rot &\text{for } mT\leq t\leq mT+t_R\\
\mathbf{\Omega}^\zer &\text{for } mT+t_R<t< (m+1)T
\end{cases},
\end{equation}
with $ \mathbf{\Omega}^\zer  = \big\{\Omega^\zer_x,0,\Delta^\zer\big\} $ and $\mathbf{\Omega}^\rot =\big\{\Omega^\rot_x,\Omega^\rot_y,\Delta^\rot\big\} $ two sets of constants. The mean-field equations of motion corresponding to a generic set of constants $ \bm{\Omega}= \set{\Omega_x,\Omega_y,\Delta}  $ are
\begin{equation}\label{eq:MFEoM}
\begin{cases}
\dot{S}^x=2\Omega_y(2n-1)-\Delta \Sy-Vn\Sy-\frac{\Gamma}{2}\Sx\\
\dot{S}^y=-2\Omega_x(2n-1)+\Delta \Sx+Vn\Sx-\frac{\Gamma}{2}\Sy\, ,\\
\dot{n}=\Omega_x\Sy-\Omega_y\Sx-\Gamma n
\end{cases}
\end{equation}
where we introduced the effective interaction coupling $V = 2 N^{-1} \sum_{k \neq p} V_{kp}$. Equation~\eqref{eq:MFEoM} can then be straightforwardly generalized to the periodically-driven case simply by updating the parameters in time according to the rules defined in \eqref{eq:omegadriving}.

The next step is to define the rotational dynamics $ \sL^\rot $ consistently with our requirements, i.e.~such that it connects the BoAs of $\bm{M}^{\mathrm{ss}}_1$ and $\bm{M}^{\mathrm{ss}}_2$. Instead of attempting to formally map out the latter two, we take a more intuitive approach: Since the stationary solutions are defined in terms of two vectors $\bm{M}^{\mathrm{ss}}_1$ and $\bm{M}^{\mathrm{ss}}_2$, we look for a global rotation $U$ exchanging their respective directions and for a regime where this is sufficient to map a stationary state into the BoA of the other. By defining the versors $\bm{m}^{\mathrm{ss}}_i = \bm{M}^{\mathrm{ss}}_i / \abs{\bm{M}^{\mathrm{ss}}_i}$, $U$ can be described as a rotation by $\pi$ around their bisecant. 
In the spin representation, $U = \exp\left[-i\frac{\pi}{2}\sum_{k}\bm{\sigma_k}\cdot\bm{d}\right]$, where $ \bm{\sigma}_k=(\sxk,\syk,\szk) $ and $ \bm{d}=(d_x,d_y,d_z) $ is defined such that $\bm{D} = \bm{m}^{\mathrm{ss}}_1 + \bm{m}^{\mathrm{ss}}_2$ and $\bm{d} = \bm{D}/ \abs{\bm{D}}$. We then choose $\mathbf{\Omega}^\rot$ in such a way that the non-interacting part of the Hamiltonian $H^\rot$ would perform precisely the rotation $U$ in a time $t_R$, namely
\begin{equation}\label{eq:rotation}
\Omega_x^\rot=\frac{\pi d_x}{2t_R},\quad\Omega_y^\rot=\frac{\pi d_y}{2t_R},\quad\Delta^\rot=\frac{\pi d_z}{t_R}.
\end{equation}
Clearly, this does not guarantee that each stationary state is mapped in the other's BoA; however, the effectiveness of this choice can be verified \emph{a posteriori} and it works for a wide range of parameter values. Notice that $t_R$ can be freely tuned to be small so that interactions and dissipation have negligible effects.

We remark that demanding each stationary state to be mapped by $\exp [\sL^\rot t_R]$ into the BoA of the other is a much looser requirement than demanding the exact mapping between the two stationary state solutions, i.e.~$\bm{M}^{\mathrm{ss}}_1 \to \bm{M}^{\mathrm{ss}}_2$ and vice versa. Hence, imperfections in the rotation procedure will not be relevant as long as its end points [squares in Fig.~\ref{fig:DTC-MF}(a)] are in the correct BoA. Indeed, the subsequent evolution, for times $t \gg \tau_3$, guarantees that the state is driven again close to the desired stationary point. This clearly adds to the robustness (or \emph{rigidity}) of the DTC phase in the proposed mechanism.

In Fig.~\ref{fig:DTC-MF}(b) we show the typical $2T$-periodic evolution of an observable in the DTC phase of the mean-field equations \eqref{eq:MFEoM}. The stationary phase diagram in the $\Omega_x^\zer-V$ plane, obtained via numerical solution of the same equations, is instead displayed in Fig.~\ref{fig:DTC-PD}(a). The colored area corresponds to the bistable region of the mean-field model, where a DTC can be constructed via our procedure. As an order parameter we consider a normalized Fourier component $ f_2 = |F(1/2)|^2/\sum_{j \in \Z} |F(j/K )|^2 $, where $ F(j/K) = (1/KT) \int_{t_w}^{t_w + KT} \rmd \tau \, S^x(\tau) \, \rme{-\frac{2\pi j}{KT} \tau} $ with the waiting time $t_w$ long enough to avoid the transient part of the dynamics, $K$ an integer $\gg 1$ and $j  \in \Z$. With our specific choice of $\sL^\rot$, DTC order is indeed displayed over a finite region of the parameter space. We also studied the robustness of the DTC phase against fluctuations of the parameters of the rotation, for instance $ \Omega_{x,y}^\rot(\varepsilon)=\Omega_{x,y}^\rot+\varepsilon \Omega_{x,y}^\rot$ with $ |\varepsilon|<1 $. The DTC remains stable over a reasonably wide range of $ \epsilon \sim 0.1$. 

\textit{Finite-size systems ---} We now turn to the case of finite-size systems to explore how the DTC phase emerges as the number of spins $N$ is increased. First, we focus our attention on a fully-connected model with $ V_{kp}=V_0/N \ \forall k,p$, which is expected to match the mean-field predictions in the thermodynamic limit. With this choice of the interactions, the model becomes permutationally symmetric~\cite{Shammah:2018,Kirton:2017} and one can study its dynamics in the totally-permutationally-symmetric subspace~\cite{perminvestatenote}. In Fig.~\ref{fig:DTC-PD}(b) we display the inverse gap $\tau_2 = 1/\abs{ \re{(\lambda_2)}  }$ in the same range of parameters used in panel (a) for a system of $N = 28$ spins. The dark zone shows a closing of the gap of $\sL^\zer$ which nicely fits with the mean-field bistable region. Within the same region, $\tau_2$ increases with $N$, whereas outside it seems to converge to a size-independent value. In the same range of parameters, $\abs{\re{(\lambda_3)}}$ does not strongly depend on $N$, leading to the emergence of a metastable regime for large enough $N$. In Fig.~\ref{fig:DTC-PD}(c) we show the stroboscopic dynamics of $S_x (t) $ (collecting data points only every period $T$) generated by the fully-connected model, where we set the parameters of $\sL^\rot$ to the mean-field ones. Here, a DTC phase emerges only at short times and eventually dies out exponentially fast. The typical lifetime of the oscillations $ \tau_{\mathrm{DTC}} $, however, increases with the system size, consistently with the expectation that the fully-connected model should reproduce the mean-field results in the thermodynamic limit.  

\begin{figure}
	\centering
	\includegraphics[width=\columnwidth]{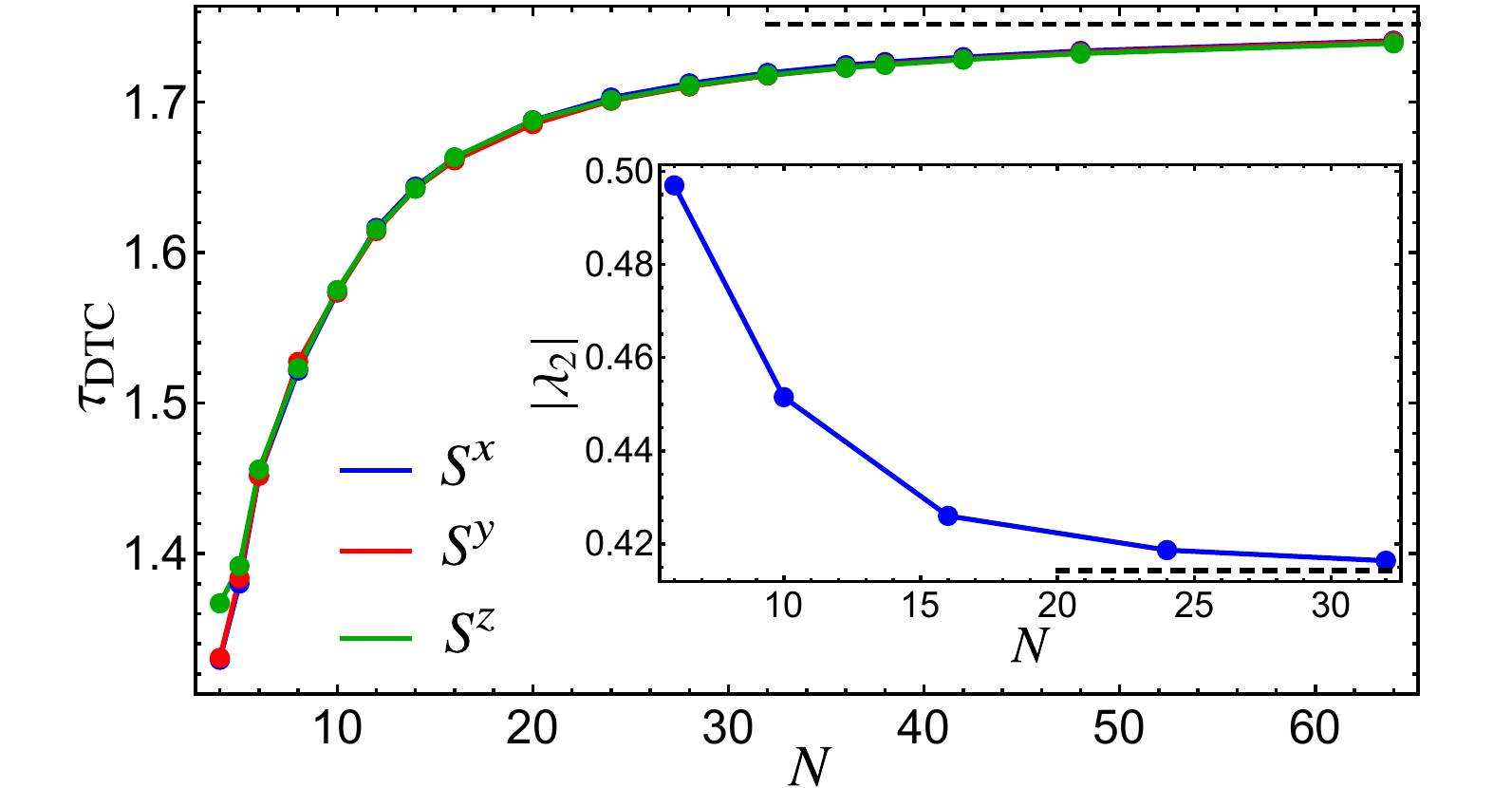}
	\caption{Lifetime $ \tau_{\mathrm{DTC}} $ of the oscillations with period $ 2T $ as a function of $ N $ in a 1D Rydberg gas with nearest-neighbor interactions extracted from the stroboscopic dynamics of $ S^{x/y/z}(t)$. The state at $ t=0 $ is with all spins pointing down in the $z$ direction, while the transformation parameters are given by Eq.~\eqref{eq:rotation} with $ d_x=0.1387 $, $ d_y=0.6824 $, and $ d_z=-0.7177 $. The black dashed line represents the asymptotic value of $ \tau_{\mathrm{DTC}} $ obtained by fitting with the function $ g(N)=a-b/N^{c} $. Inset: gap $ |\lambda_2| $ as a function of $ N $ associated with the $ \sL^\zer $ dynamics of the main panel. Data points are obtained by fitting the long-time decay of $ S^x(t) $ with an exponential decay $ \propto e^{-\lambda_2 t} $. The black dashed line is the asymptotic value obtained by a fit with $ g(N)$ as before. Here, $ \Omega^\zer_x=\Gamma $, $ V_0=1.6\ \Gamma $, $ \Delta^\zer=-3.5\ \Gamma $, $ T=2\, \Gamma^{-1} $, and $ t_R=10^{-2}\, \Gamma^{-1} $. }
	\label{fig:tau-PV}
\end{figure}

Finally, we briefly turn to a finite-dimensional case. Since the phase transition in the model we have considered has a lower critical dimension of $2$~\cite{Hu:2013,Marcuzzi:2014}, it would be clearly ideal to simulate a large two-dimensional lattice. This is beyond current numerical capabilities. Hence, we focus instead on an accessory feature of our mechanism, i.e., that there should be a correlation between the spectral gap $ |\lambda_2| $ and the lifetime of the DTC, $ \tau_{\mathrm{DTC}} $, which is in principle observable in one dimension as well. In Fig.~\ref{fig:tau-PV} we report results for a one-dimensional nearest-neighbor case ($ V_{kp}=V_0\delta_{p,k\pm1} $), which can be efficiently investigated by employing a time-evolving block-decimation algorithm~\cite{Vidal:2003,Prosen:2009,Pfeifer:2014ncon}. 
As in the previous case, the stroboscopic dynamics of a typical observable displays an oscillatory behavior with period $ 2T $ and amplitude decaying exponentially in time. 
The comparison between $ \tau_{\mathrm{DTC}} $ and $ |\lambda_2| $ up to $ N=32 $ highlights
the emergence of the expected correlation in a short-range system: The smaller the gap becomes, the longer the DTC structure survives; at the scale where one saturates to a finite value, the other saturates as well. This supports the conjecture that metastable open quantum systems with a closing gap develop, under appropriate periodic driving, DTC phases also in low dimensions.
	
\textit{Conclusions ---} We have discussed a general mechanism for engineering a DTC in driven open quantum systems subject to metastability. This requires neither disorder nor explicit symmetries, although the phase transition associated to the closing of the gap may display an emergent one. We have shown the emergence of a DTC order in a specific case taken from the physics of dissipative Rydberg gases. This, in turn, means that Rydberg systems may represent in the future an interesting platform for the study of dissipative DTC phases. 

\begin{acknowledgments}
The research leading to these results has received funding from the  European  Research  Council  under  the  European Unions Seventh Framework Programme (FP/2007-2013)/ERC Grant Agreement No.~335266 (ESCQUMA) and the EPSRC Projects No.~EP/M014266/1, EP/R04340X/1 and EP/N03404X/1. I.L.~gratefully acknowledges funding through the Royal Society Wolfson Research Merit Award.
\end{acknowledgments}

% Footnotes
\footnotetext[1]{More precisely, the mean-field analysis suggests the universality class of the transition to be the so-called \emph{model A} \cite{Hohenberg:1977}}
\footnotetext[2]{See Supplemental Material for further details, which includes Ref.~\cite{Beterov:2009}. }

% Bibliography
\bibliography{DTCbibliography}

\begin{thebibliography}{81}
\bibitem [{per()}]{perminvestatenote}%
  \BibitemOpen
  \href@noop {} {}\bibinfo {note} {Note that this stricly holds if and only if
  the system is initialized at $t=0$ in a permutationally invariant
  state.}\BibitemShut {Stop}%
\end{thebibliography}

\pagebreak

\widetext
\begin{center}
	\textbf{\large Supplemental Material for "Discrete time crystals in the absence of manifest symmetries or disorder in open quantum systems"}
\end{center}

%%%%%%%%%% Merge with supplemental materials %%%%%%%%%%
%%%%%%%%%% Prefix a "S" to all equations, figures, tables and reset the counter %%%%%%%%%%
\setcounter{equation}{0}
\setcounter{figure}{0}
\setcounter{table}{0}
\makeatletter
\renewcommand{\theequation}{S\arabic{equation}}
\renewcommand{\thefigure}{S\arabic{figure}}

\renewcommand{\bibnumfmt}[1]{[S#1]}

%%%%%%%%%% Prefix a "S" to all equations, figures, tables and reset the counter %%%%%%%%%%
\par
\begingroup
\leftskip2cm
\rightskip\leftskip
\small In this Supplemental Material we provide some additional details on the dissipative Rydberg model we have considered in the main text to illustrate the working principles of the general mechanism we have proposed to engineer discrete time crystals in open quantum systems. 
\par
\endgroup

\section{Dissipative Rydberg model} 
\begin{figure}
	\centering
	\topinset{(a)}{	\raisebox{-0.5\height}{\includegraphics[height=0.13\textheight]{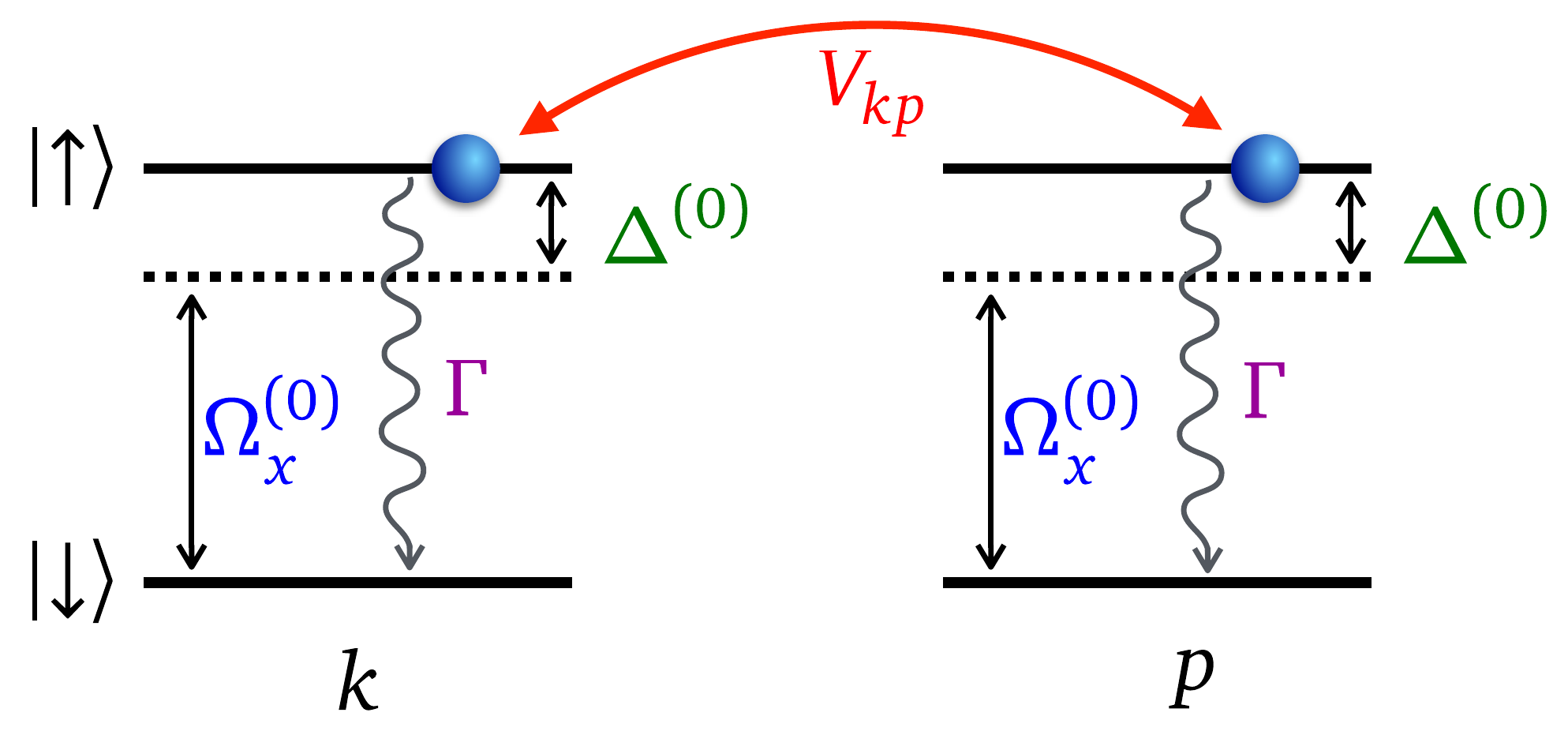}}}{-0.6 cm}{0.0 cm}\qquad
	\topinset{(b)}{	\raisebox{-0.5\height}{\includegraphics[height=0.19\textheight]{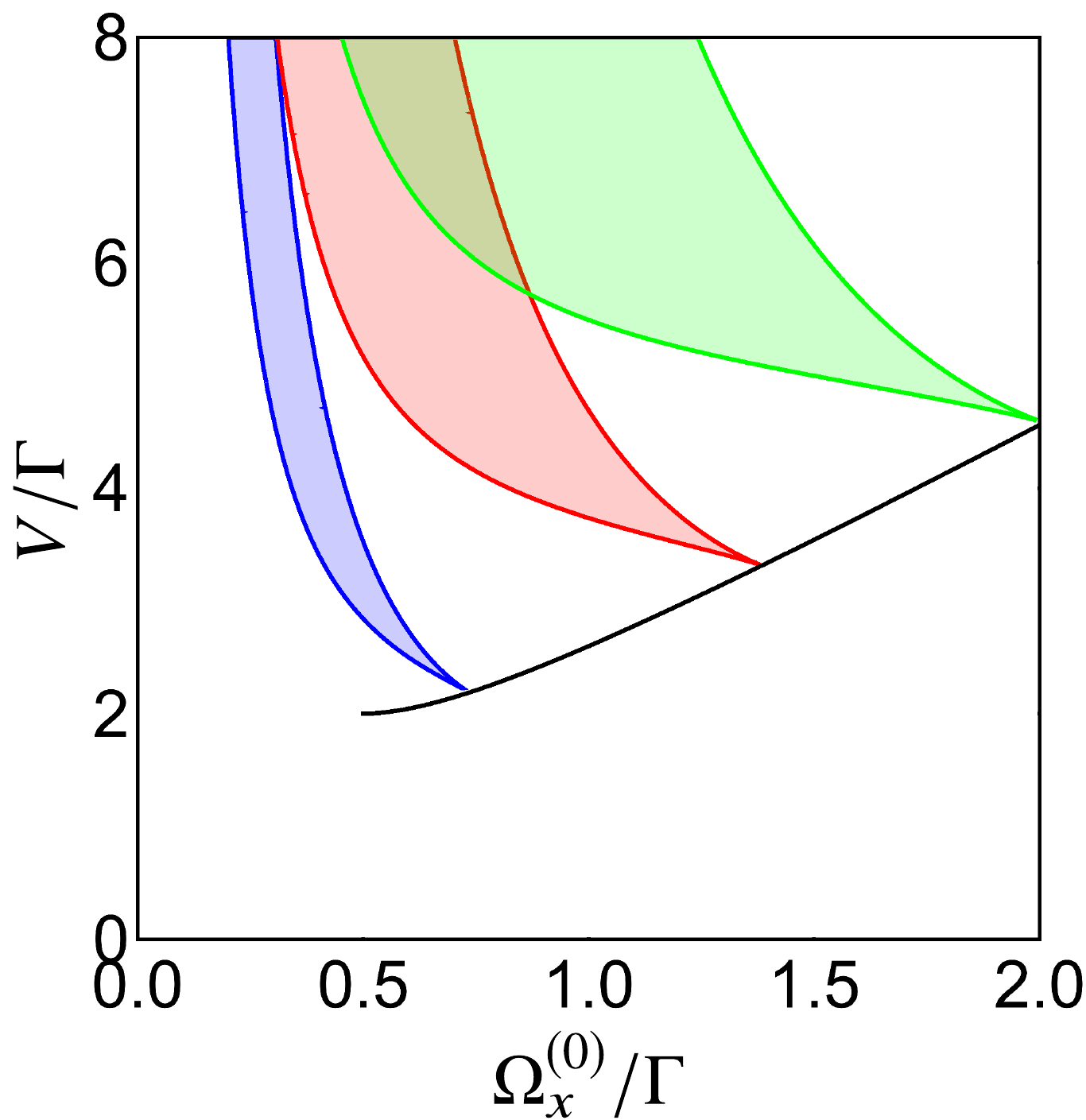}}}{0.1 cm}{-0.2 cm}\qquad
	\topinset{(c)}{	\raisebox{-0.5\height}{\includegraphics[height=0.175\textheight]{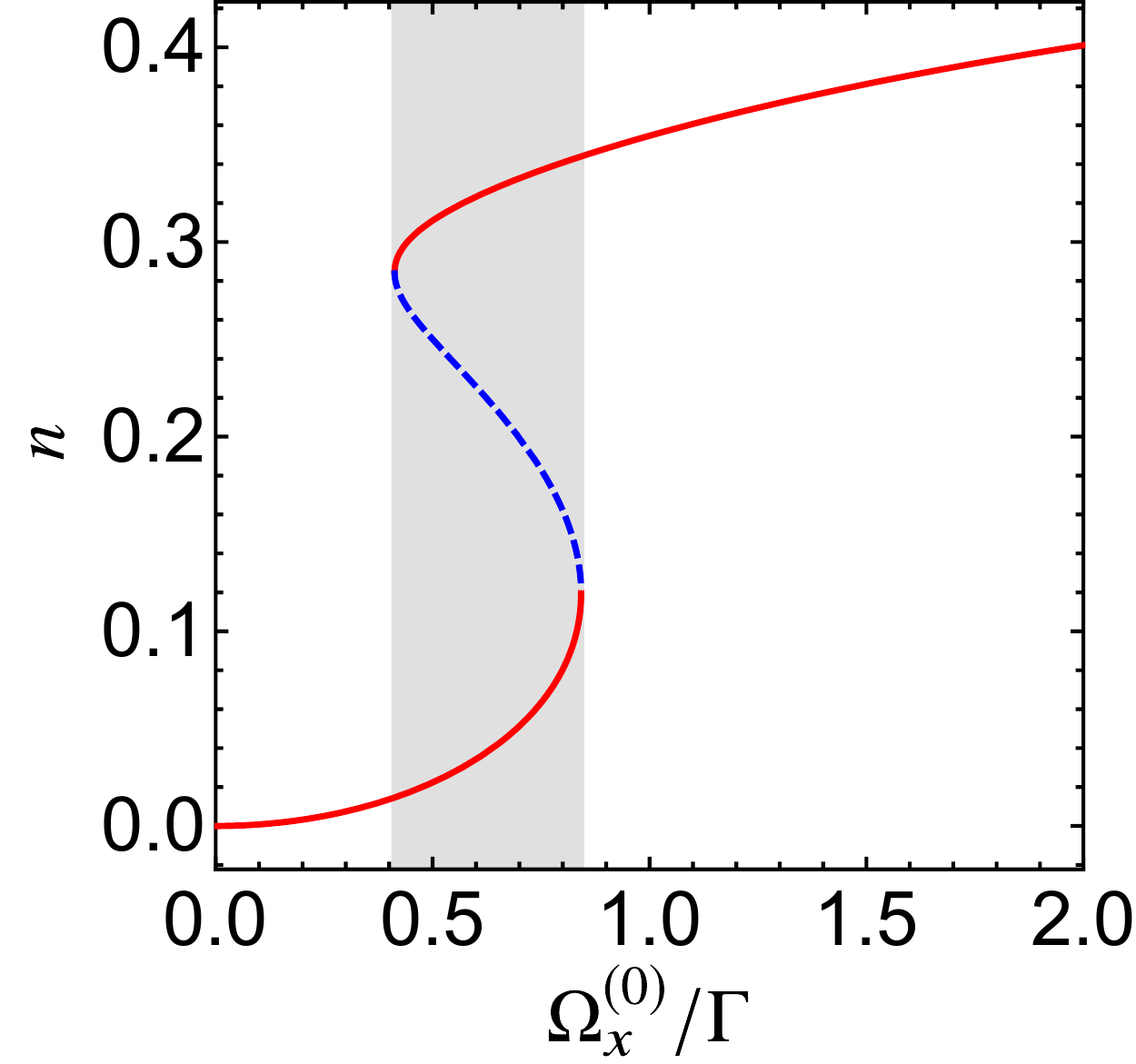}}}{-0 cm}{-0.1 cm}
	\caption{In panel (a) we sketch the energy levels of two atoms of the dissipative Rydberg model, which are laser-coupled with a laser with Rabi frequency $ \Omega_x^{(0)} $ and detuning $ \Delta^{(0)} $. Two atoms, at sites $ k $ and $ p $, which occupy simultaneously the Rydberg state interact with the interaction potential $ V_{kp} $. Once excited to the Rydberg state, an atom will decay to the ground state due to spontaneous emission at rate $ \Gamma $. Panel (b) shows the mean-field phase diagram in the $ \Omega_x^{(0)}-V $ plane for $ \Delta^{(0)}=-2\ \Gamma $ (blue), $ \Delta^{(0)}=-3.5\ \Gamma $ (red), and $ \Delta^{(0)}=-5\ \Gamma $ (green). Colored regions denote the presence of two stable asymptotic solutions to Eq.~\eqref{SM:eq:MFEoM}. The black line shows the locus of the critical points $ (\Omega_{x,c}^{(0)},V_{c}) $ as $ \Delta^{(0)} $ is varied continuously: Note that, in order for a bistable regime to emerge, a finite interaction potential $ V $ is required. In panel (c) the behavior of the Rydberg occupation number $ n $ as a function of $ \Omega^{(0)}_x $ is shown. Here, $ \Delta^{(0)}=-3.5\ \Gamma $ and $ V=6\ \Gamma $. Inside the shaded region, corresponding to the emergence of a bistable regime, Eq.~\eqref{SM:eq:n} admits three real solutions, two of which are stable (red, smooth lines) and one unstable (blue, dashed line). }
	\label{SM:fig:dRg}
\end{figure}

In the main text we have focused on a dissipative system of $ N $ Rydberg atoms. Here, the ground state of each atom is laser-coupled to high-energy (Rydberg) level with large principal quantum number $ n $. Since dipole-dipole interactions are $ \propto n^{11} $, when two atoms are both excited to the Rydberg level they exhibit a strong and long-range interaction~\cite{Low:2012,Gallagher:2005,Saffman:2010}. In particular, the interaction potential is $ V(R)\sim R^{-\alpha} $, being $ R $ the distance between the excited atoms, with either $ \alpha=3 $ (dipole-dipole interactions) or  $ \alpha=6 $ (van der Waals interactions) depending on the particular choice of the Rydberg state. Once excited, an atom will decay from the Rydberg to the ground state due to spontaneous emission and/or blackbody radiation. While the latter, which would couple the Rydberg state to nearby levels, can be neglected at low temperature~\cite{Beterov:2009}, spontaneous emission to the ground state can directly connect the Rydberg level to the ground state or it can pass through intermediate low-lying energy levels. However, since the lifetime of the latter is, in general, much shorter than the Rydberg state one, the whole spontaneous emission process can be modeled as a direct transition from the Rydberg to the ground state at rate $ \Gamma $. Therefore, assuming that the atoms are arranged in a regular lattice with fixed lattice constant, each atom can be modeled as an effective two-level system, which can thus be described in terms of fictitious spin-$ \frac{1}{2} $ particles. The two energy levels are coupled by a laser with Rabi frequency $ \Omega^{(0)}_x $ and detuning $ \Delta^{(0)} $ with respect to the ground state-Rydberg state transition. We denote with $ \ket{\downarrow} $ and $ \ket{\uparrow} $ the ground state and the Rydberg state, respectively. The system is sketched in Fig.~\ref{SM:fig:dRg}(a). 
Within the rotating wave approximation~\cite{Barnett:2005}, the Hamiltonian of the dissipative Rydberg model is~\cite{Lee:2011,Lee:2012,Ates:2012,Low:2012,Marcuzzi:2014}
\begin{equation}
	H^{(0)}=\sum_{k=1}^{N}\left[\Omega_x^{(0)} \sxk+\Delta^{(0)}  n_k\right]+\sum_{k\neq p}^N V_{kp}n_k n_p
\end{equation}
Here, $k$ and $p$ are site indices, $\sigma_k^{\mu}\ (\mu = x,y,z)$ denote the Pauli matrices acting on the $k-$th spin, $ n_k=(\mathbb{I}_k+\szk)/2 $ is the $k-$th site number operator of the Rydberg level, and $V_{kp}$  describes two-body interactions between the $ k- $th and $ p- $th atoms, respectively. The finite lifetime of the Rydberg state due to spontaneous emission is taken into account through the dissipative term 
\begin{equation}\label{SM:eq:D}
	\sD[\rho]= \Gamma \sum_{k=1}^{N}\left[\smmk \rho \sppk- \frac{1}{2}\left\{\sppk \smmk ,\rho\right\}\right],
\end{equation} where $\sigma_k^{\pm} = (\sxk-i \syk)/2$. Therefore, the time-evolution of the density matrix of the model, $ \rho $, is governed by the quantum master equation (QME)~\cite{BreuerPetruccione}
\begin{equation}\label{SM:eq:QME}
	\partial_t\rho=\mathcal{L}^{(0)}[\rho]=-i[H^{(0)},\rho]+\sD[\rho].
\end{equation}
We now summarize the main results of the mean-field analysis reported in Ref.~\cite{Marcuzzi:2014}. Focusing on the expectation values of the single-particle observables $S^\mu = \av{\sigma_k^\mu}$  and $n = \av{n_k}$, where the site index $k$ has been dropped assuming translational invariance, one gets the following mean-field equations of motion
\begin{equation}\label{SM:eq:MFEoM}
	\begin{cases}
		\dot{S}^x=-\Delta \Sy-Vn\Sy-\frac{\Gamma}{2}\Sx\\
		\dot{S}^y=-2\Omega_x(2n-1)+\Delta \Sx+Vn\Sx-\frac{\Gamma}{2}\Sy,\\
		\dot{n}=\Omega_x\Sy-\Gamma n
	\end{cases}
\end{equation}
where $V = 2 N^{-1} \sum_{k \neq p} V_{kp}$ is the mean-field effective interaction potential. The stationary solutions of the above equations can be found by setting to zero the right-hand side of Eq.~\eqref{SM:eq:MFEoM}, i.e., $ \dot{S}^x=\dot{S}^y=\dot{n}=0 $. By doing so, one obtains the following algebraic equation for $ n $
\begin{equation}\label{SM:eq:n}
	n\left[2+\frac{1}{4}\left(\frac{V}{\Omega^{(0)}_x}\right)^2+\left(\frac{V}{\Omega^{(0)}_x}\right)^2\left(\frac{\Delta^{(0)}}{V}+n\right)^2\right]=1.
\end{equation} 
Being a cubic polynomial with real coefficients, Eq.~\eqref{SM:eq:n} admits either one or three real solutions, depending on the values of $ \Omega^{(0)}_x $, $ \Delta^{(0)} $, and $ V $. Panel (b) of Fig.~\ref{SM:fig:dRg} shows the phase diagram of the dissipative Rydberg model for different fixed values of $ \Delta^{(0)} $. Here, shaded areas correspond to the emergence of a bistable regime, associated with the coexistence of two real stable solutions of Eq.~\eqref{SM:eq:n}, $ \bm{M}^{\mathrm{ss}}_i=(S^x_i,S^y_i,S^z_i)\ (i=1,2) $. Note that, if present, the mean-field bistable region is enclosed by two spinodal lines which coalesce with zero net angle into the critical point~\cite{Marcuzzi:2014} 
\begin{equation}\label{SM:eq:CP}
	\begin{cases}
		\Omega_{x,c}^\zer =\sqrt{\frac{1}{2}\left[\frac{(\Delta^\zer)^2}{3}-\frac{\Gamma^2}{4}\right]},\\
		V_c=-\frac{8}{27}\frac{(\Delta^\zer)^3}{(\Omega_{x,c}^\zer)^2}.
	\end{cases}
\end{equation}
By inspecting the behavior of the system near this critical point, it is finally possible to show that a symmetry line, passing through $ (\Omega^{(0)}_{x,c},V_c) $ and along which the solutions to Eq.~\eqref{SM:eq:MFEoM} exhibit a $ \mathbb{Z}_2 $ invariance, emerges. The two stable stationary solutions, $ \bm{M}^{\mathrm{ss}}_1 $ and $ \bm{M}^{\mathrm{ss}}_2 $, are related by this emergent $ \mathbb{Z}_2 $ symmetry which, within the bistable region, is thus spontaneously broken. The interested reader is referred to Ref.~\cite{Marcuzzi:2014} for further details. 

% Footnotes
%\footnotetext[1]{Note...}

\end{document}